\PassOptionsToPackage{pdfpagelabels=false}{hyperref} 
\documentclass[a4paper,fleqn,usenatbib]{mnras}

\usepackage{newtxtext,newtxmath}

\usepackage[T1]{fontenc}
\usepackage{ae,aecompl}

\usepackage{graphicx}	
\usepackage{amsmath}	
\usepackage{amssymb}	

\title[Probing the formation of the NSC with pulsars]{Probing the formation history of the nuclear star cluster at the Galactic Centre with millisecond pulsars}

\author[Abbate et al.]{
F. Abbate$^{1,2}$,\thanks{E-mail: f.abbate@campus.unimib.it}
A. Mastrobuono-Battisti$^{3}$,
M. Colpi$^{1,4}$,
A. Possenti$^{2}$,
A. C. Sippel$^{3}$ and \newauthor
M. Dotti$^{1,4}$ 
\\
$^{1}$ Departement of Physics G. Occhialini, University of Milano Bicocca, Piazza della Scienza 3, 20126 Milano, Italy\\
$^{2}$ INAF - Osservatorio Astronomico di Cagliari, Via della Scienza 5, I-09047 Selargius (CA), Italy\\
$^{3}$ Max-Planck Instiut f\"ur Astronomie, K\"onigstuhl 17, D-69117 Heidelberg, Germany\\
$^{4}$ Istituto Nazionale di Fisica Nucleare, INFN, Milano Bicocca, Piazza della Scienza 3, 20126 Milano, Italy\\
}

\date{Accepted XXX. Received YYY; in original form ZZZ}

\pubyear{2017}

\begin{document}
\label{firstpage}
\pagerange{\pageref{firstpage}--\pageref{lastpage}}
\maketitle

\begin{abstract}
The origin of the Nuclear Star Cluster in the centre of our Galaxy is still unknown. One possibility is that it formed after the disruption of stellar clusters that spiralled into the Galactic Centre due to dynamical friction. 
We trace the formation of the Nuclear Star Cluster around the central black hole, using state-of-the-art $N$-body simulations,  and follow the 
dynamics of the neutron stars born in the clusters. We then estimate the number of Millisecond Pulsars (MSPs) that are released in the Nuclear Star Cluster during its formation. 
The assembly and tidal dismemberment of globular clusters lead to a population of MSPs distributed 
over a radius of about 20 pc, with a peak near 3 pc. No clustering  is found on the sub-parsec scale. 
We simulate the detectability of this population with future radio telescopes like the MeerKAT radio telescope and SKA1, and find that about of order ten MSPs can be observed over this large volume, with a paucity of MSPs within the central parsec.  This helps discriminating this scenario from the in-situ formation model for the Nuclear Star Cluster that would predict an over abundance of MSPs closer to the black hole.
We then discuss the potential contribution of our MSP population  to the gamma-ray excess at the Galactic Centre.
\end{abstract}

\begin{keywords}
Galaxy: centre -- globular clusters -- pulsars -- Galaxy: formation
\end{keywords}



\section{Introduction}

Nuclear star clusters (NSCs) are dense stellar systems observed in a large fraction of galactic nuclei \citep{Carollo1998, Boker2002}.  They have masses between $10^6M_\odot$ and  $10^7M_\odot$, effective radii of less than $5$ pc \citep{Boker2004, Cote2006, Georgiev2014} and often host a supermassive black hole (SMBH) at their centre \citep{Georgiev2016}. 

The central parsecs of our Galaxy host a NSC \citep{Genzel2010,Mapelli2015} with half-light radius of $4.2 \pm 0.4$ pc \citep{Schodel2014}, a mass of $2-3 \times 10^7 \,\rm M_{\odot}$ \citep{Feldmeier2014, Schodel2014} and a central black hole of  $4.30 \pm 0.20_{\rm stat} \pm 0.30_{\rm sys}\times 10^6\,\rm M_\odot$ \citep{Gillessen2009}. 

The formation mechanism of NSCs is not yet known and two are the main scenarios proposed so far: (1) the \emph{in-situ} formation scenario  where infall of gas to the centre of the galaxy and subsequent star formation are responsible for the formation of the cluster and its stellar populations  \citep{Loose1982, Milosavljevic2004, McLaughlin2006, Aharon2015}; (2) the {\it cluster-inspiral} scenario where stellar clusters spiral into the galactic centre under the action of dynamical friction and slowly accumulate to build up the NSC \citep{Tremaine1975,Capuzzo-Dolcetta1993, Antonini2012, Gnedin2014, Perets2014, Arca-Sedda2014,Arca-Sedda2015, Antonini2015, Arca-Sedda2017b, Arca-Sedda2017}. While the first scenario is dissipative  and relies on violent episodic events of gas inflows, the second is dissipationless and regular over the entire lifespan of the NSC.

These two formation processes may not be mutually exclusive, and can work in concert \citep{Antonini2015}.
In particular, the {\it cluster-inspiral} scenario can account for both the morphological and kinematic properties of the NSC at the Galactic centre \citep{Tsatsi2017}. On the other hand the young massive stars in the Galactic NSC \citep{Schodel2003, Ghez2005, Genzel2010} show the presence of an ongoing star formation.

The central regions of the Galaxy are thought to host a large population of both ordinary pulsars and millisecond pulsars \citep[MSPs,][]{Pfahl2004, Wharton2012}. 
The large number of massive and young stars suggests that an abundant population of ordinary pulsars could stem from the ongoing star-formation. Moreover, the Galactic Centre hosts an excess of X-ray sources comparable to the excess measured in globular clusters \citep{Muno2005, Haggard2017}. This might be indicative of a large population of MSPs, similar to the ones observed in globular clusters. 
This population was for long thought to be unobservable with current telescopes because of the very large expected gas densities in the central regions which would create strong scattering screens \citep{Lazio1998}. The scattering screens would induce a temporal smearing which would cover completely the pulsed emission of the pulsars.

The discovery of the Galactic Centre magnetar SGR J1745-29 at just 0.1 pc from the central SMBH, Sgr A* indicated that the scattering screen could be up to three orders of magnitude weaker than what expected \citep{Spitler2014}, at least along some favourable lines of sight. Moreover, the angular broadening scale of the magnetar is in excellent agreement with the one measured for Sgr A*, suggesting that the two sources lie behind the same scattering screen \citep{Bower2014}. The angular broadening scale is also similar to what measured in masers at $\sim 1-50$ pc from the Galactic Centre. Therefore we can argue that the effects of the scattering screen are similar to those seen in SGR J1745$-$29 over a significant portion of the Galactic Center region.

In spite of these predictions, the numerous surveys  -- some of them performed also at relatively high radio frequencies, from 3 GHz to 15 GHz -- focused on finding this population of ordinary and millisecond pulsars in the central parsecs of the Galactic region \citep{Johnston2006, Deneva2009, Macquart2010, Bates2011} have not detected a single pulsar. This disagreement between predictions and observations is known as \emph{The Missing Pulsar Problem} \citep{Dexter2014}.

In fact, the Galactic Centre pulsar population might also be responsible for the excess that the \emph{Fermi} satellite detected in gamma-ray \citep{Hooper2011, Gordon2013,Daylan2016}. This excess peaks at $\sim 2$ GeV, is roughly spherical and extends to $\sim 10^{\circ}-20^{\circ}$ (1.5-3 kpc) from the Galactic Centre. The most probable sources of the excess are the annihilation from dark matter particles \citep{Hooper2011, Daylan2016}, the emission from diffuse cosmic rays \citep{Gaggero2017} and the emission from a population of MSPs \citep{Abazajian2011, Gordon2013}. Wharton et al. (\citeyear{Wharton2012}) estimated that to produce the observed emission the number of MSPs within 1 pc should be $\lesssim 5000$.  
\cite{Brandt2015} suggested, and \cite{Arca-Sedda2017c} confirmed, that the emission from MSPs from disrupted globular clusters could explain the observed excess. 
Other authors instead claim that this mechanism could only explain a few percent of the total excess \citep{HooperMohlabeng2016, HooperLinden2016, Haggard2017}.

Globular clusters are known to be breeding grounds for the formation of  MSPs \citep{Camilo2005, Freire2013}. 
Thus, in the cluster-inspiral scenario globular clusters are expected to deposit their population of MSPs which are then inherited by the NSC.

In the \emph{in-situ} formation scenario, ordinary pulsars form out of regular star formation and should be found where young massive stars are observed, $\sim 0.5$ pc \citep{Bartko2009,Yusef-Zadeh2013, Mapelli2015}. Field binaries and binaries that form via dynamical interactions in the densest regions of the NSC can produce a population of MSPs \citep{Faucher-Giguere2011}.

Since ordinary pulsars live only for a few tens of Myr, the current population can not carry any information on the formation scenario. By contrast, MSPs  are expected to live and emit radio pulsations for 1-10 Gyr timescales and can be used to disentangle the two formation processes.

This paper is a first attempt to probe the cluster-inspiral scenario using MSPs as tracers of the dynamical formation of the Galactic NSC.
The paper is organised as follows:
in $\S$ \ref{Cluster_infall_scenario} we describe the $N$-body simulations of the formation of the NSC through the infall of 12 globular clusters on the massive black hole at the centre of the Milky Way; in $\S$ \ref{NSglobular_cluster} we infer the spatial distribution of neutron stars  and 
their number from an independent {\it N}-body simulation of a less massive globular cluster, whose dynamical evolution is simulated with full stellar evolution; in $\S$ \ref{NSgalactic_center} we scale this distribution to describe neutron stars  in the globular clusters of the NSC simulations, taking into account the different masses, radii, and evolutionary states.  We tag and follow the neutron stars during the formation and evolution of the NSC; in $\S$ \ref{Rec_fraction} we estimate how many neutron stars could have been recycled as MSPs. In this way, we are able to reconstruct the spatial distribution of MSPs in the Galactic Centre and study its properties. 
Once we have this distribution, we focus on the detectability of this MSP population with future generation radio telescopes in $\S$ \ref{observability} and we show our results in $\S$ \ref{results}. In $\S$ \ref{discussion} we briefly discuss the possible detections in the in-situ formation mechanism and we also test the consistency of our results with the gamma-ray emission observed by the {\it Fermi} satellite in this region.

\section{Neutron stars distribution in cluster-inspiral scenario}

\subsection{Globular cluster-inspiral model}\label{Cluster_infall_scenario}

The cluster-inspiral scenario that we refer to in this paper is described in Antonini et al. (\citeyear{Antonini2012}), Perets and Mastrobuono-Battisti (\citeyear{Perets2014}) and Tsatsi et al. (\citeyear{Tsatsi2017}). In these investigations,  NSCs  form via repeated mergers of dense and massive stellar clusters, which are
sinking toward the centre of a Milky Way-like bulge.

In the {\it N}-body simulation by Tsatsi et al. (\citeyear{Tsatsi2017}) used in this investigation, a NSC grows by letting 12 identical stellar clusters fall on the Galactic Centre, one every $\sim 0.85$ Gyr. The stellar clusters have a mass of $1.1 \times 10^6 M_{\odot}$ and are represented by a tidally truncated King model \citep{King1966} with core radius of $\sim 0.5$ pc, half mass radius of $\sim 1.2$ pc and the dimensionless King potential $W_0$ is 5.8. The half mass relaxation time for these clusters is $\sim 0.3$ Gyr.
The Galactic Centre is composed of a nuclear disc of mass $M_{\rm disk}=10^8 \, \rm M_{\odot}$ and a central massive black hole of mass $4 \times 10^6 \,\rm M_{\odot}$.  
 The black hole gravitational field acts as attractor and thus contributes to the building up of the NSC.
 The stellar clusters are released at a distance of $20$ pc from the Galactic Centre. The inspiral happens on a timescale of a few tens of Myr, much shorter than the half mass relaxation time. Therefore the internal evolution of the globular clusters can be neglected during the inspiral phase. 
 After the the arrival of the last globular cluster, the system is evolved for 2.2 Gyr, reaching a total simulation time of $\sim 12.4$ Gyr. Following Tsatsi et al. (\citeyear{Tsatsi2017}), we  analyse three simulations with different initial conditions on the orbital parameters. In I and II, the longitude of the ascending node $\Omega$ and inclination $i$ of the orbits are randomly chosen. In simulation III,  the constraint on $i<90^{\circ}$ allows
 to describe the sinking of  clusters formed in the central molecular zone of the Milky Way having random offsets with respect to the Galactic plane, but all sharing prograde orbits.

\subsection{Neutron stars in globular clusters} \label{NSglobular_cluster}

\begin{figure}
\begin{center}
\includegraphics[width=\columnwidth]{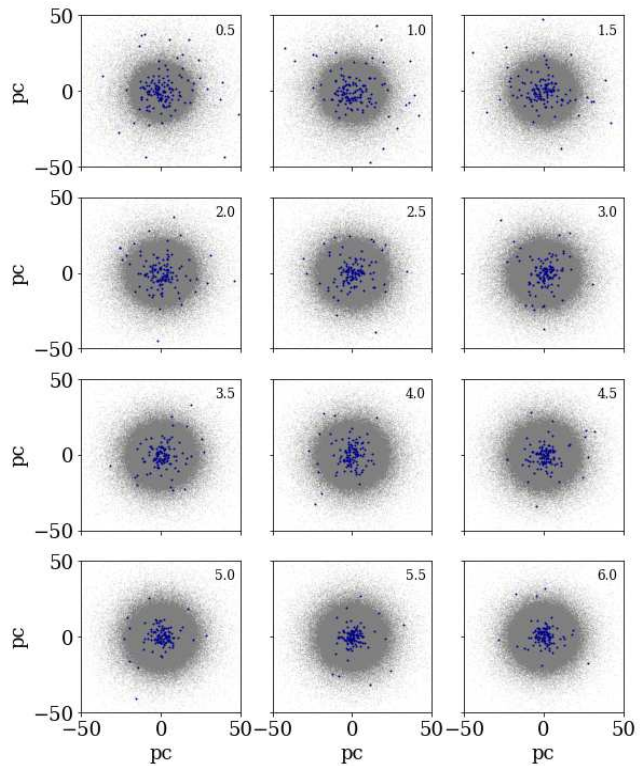}
\caption{Snapshots of the spatial distribution of the neutron stars (blue dots) in the simulated reference cluster by  \protect\cite{Sippel2013} on top the background stellar component (grey dots). The snapshots are taken one every 0.5 half mass relaxation time. The time at which each snapshot is taken is shown in each panel in units of the half mass relaxation time.}\label{NS_Sippel}
\end{center}
\end{figure}

Stellar clusters host a large number of neutron stars  that form after a few million year from the core collapse of stars with masses between $9-20\,\rm M_{\odot}$.  In order to determine their spatial distribution and density in every simulated cluster, we focus on the outcome of an $N$-body simulation of a "reference" star cluster  \citep{Sippel2013}. The cluster is evolved with {\small NBODY6} \citep{Aarseth1999,Aarseth2003,Nitadori2012} and has a total initial mass of $1.6 \times 10^5 M_{\odot}$, an initial core radius of $\sim 4$ pc, an initial half-mass radius of 6.2 pc, and is described by a Plummer density profile. The stars are initially distributed according to the initial mass function of \cite{Kroupa1993} between 0.1 $M_{\odot}$ and 50 $M_{\odot}$. The total number of stars simulated is 262 500 and are evolved accounting for stellar evolution for 12 Gyr as modelled in \citep{Hurley2000}.
In \cite{Sippel2013}, the neutron stars receive kicks at birth whose strength is adjusted in order to obtain a retention fraction of $\sim 10\%$ as is suggested by observations \citep{Pfahl2002, Pfahl2003}. 
The simulation is sampled every Gyr, which corresponds to about 0.5 half-mass relaxation times, in order to reconstruct the mass segregation of the neutron stars. The evolution is followed up to 6 half-mass relaxation times. During the evolution of the cluster the half-mass relaxation time remains constant.
The positions of the neutron stars  as a function of time is shown in Figure \ref{NS_Sippel}.

\begin{figure}
\begin{center}
\includegraphics[width=\columnwidth]{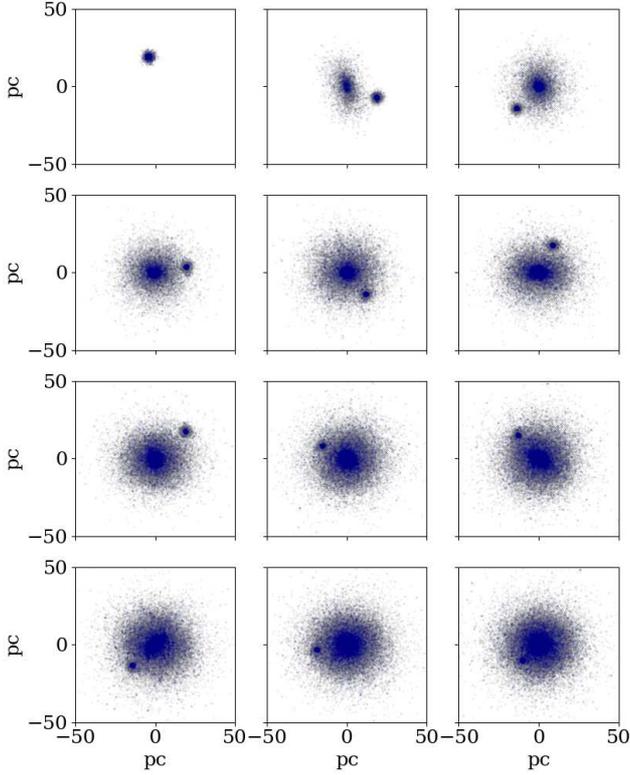}
\caption{Snapshots of the spatial distribution of the neutron stars (blue dots) in the NSC on top the background stellar component (grey dots). Each snapshot is taken when a new globular cluster is injected in the simulation.}\label{NS_NSC}
\end{center}
\end{figure}

\begin{figure}
\begin{center}
\includegraphics[width=\columnwidth]{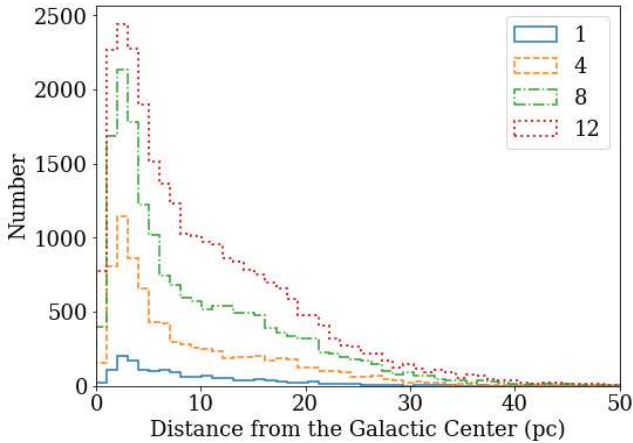}
\caption{Radial distribution of the neutron stars deposited in the NSC during the simulation. The distributions are sampled after the first, forth, eighth  and twelfth globular cluster are injected and left to evolve for 0.85 Gyrs. The number of the last globular clusters injected for each distribution is shown in the label. Each bin has a width of 1 pc. 
}\label{NS_distr_infall}
\end{center}
\end{figure}

The next step consists in rescaling and adapting the outcome of this simulation to our large scale simulation in order to track the
position of the neutron stars in each cluster during the assembly of the NSC. To set this correspondence, we divide the stars in radial bins expressed in units of the half-mass radius and measure the mass ratio of neutron stars to normal stars in every snapshot.  
After obtaining the neutron star mass fraction in every normalised bin, we mark the same fraction in the corresponding stellar clusters used in the simulation of the cluster-inspiral formation scenario. Since every cluster is injected at a different time, the spatial distribution of the neutron stars is modified by mass segregation.  The match in time is carried on attributing different spatial distributions
according to the time of cluster injection,  measured in units of the half-mass relaxation time. 
Having established these correspondences, we then follow the dismemberment of the host clusters to infer the final position of the neutron star population, at the end of the simulation. The positions and radial distribution of the neutron stars during the simulation of the formation of the NSC are shown in Figure \ref{NS_NSC} and in Figure \ref{NS_distr_infall}. The neutron stars do not cluster in the central regions of the NSC but are scattered throughout a region of 40-50 pc. As these simulations are performed with single mass particles we cannot follow the effect of mass segregation in the NSC.

\subsection{Neutron stars deposited in the Galactic Centre} \label{NSgalactic_center}

\begin{figure}
\begin{center}
\includegraphics[width=\columnwidth]{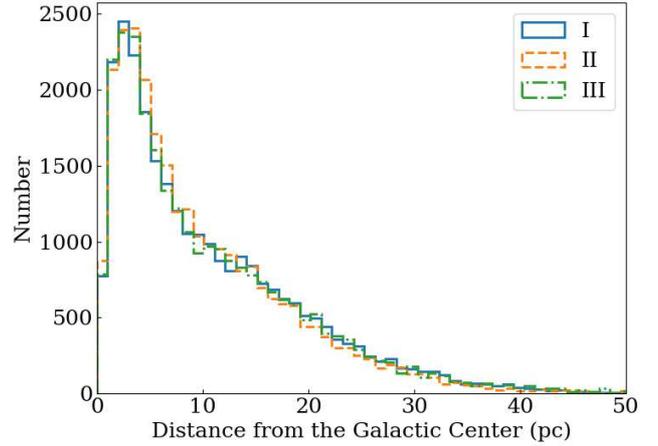}
\caption{Distribution of the neutron stars in the Galactic Centre according to the three simulations. The plots in different colours refer to the three simulations. The distributions are all compatible with one another. Each bin has a width of 1 pc and the distribution peaks at 3 pc. }\label{distribution_together}
\end{center}
\end{figure}

The radial distributions of the neutron stars deposited by the clusters at the Galactic Centre are shown in Figure \ref {distribution_together}, for the three different simulations. 
 The neutron stars  do not cluster within the central parsec. Instead, they are spread over a rather wide volume, with $\sim 87 \%$ being inside a radius of 20 pc. The peak of the distribution is at 3 pc, close to the gravitational influence radius of the central massive black hole \citep{Alexander2005, Merritt2010, Feldmeier2014, Chatzopoulos2015}. The first clusters are tidally disrupted closer to the black hole, whereas those sinking at later times have a progressively larger disruption radius, due to the build up of the NSC.  This is the cause of the broad distribution of the neutron stars, in the simulations \citep{Perets2014}.
The spatial distribution of neutron stars is weakly dependent on the initial conditions of the cluster orbits as is shown in Figure \ref{distribution_together} where results from the three simulations are plotted.  Therefore in our analysis we will use only the simulation I.

\subsection{Recycling fraction}\label{Rec_fraction}
So far, the neutron stars of our simulations have been treated as point particles.
We lacked  in tracing their small-scale dynamics in the cluster and in the NSC leading to their pairing in 
binaries, and their close interaction with companion stars, which leads to their recycling as MSPs \citep{vandenHeuvel2009}. Thus, in determining the recycling fraction our approach is necessarily statistical, and is based
on current observations of MSPs in the field and in globular clusters, and on a comparison with ordinary pulsars. We further note that our simulations do not evolve the stellar population during the NSC growth.

In the Galactic disk (within 3 kpc from the Sun) the ratio of the birthrates of MSPs to ordinary pulsars is found to be $\geq 10^{-3}$ \citep{Lyne1998}. Using this result Wharton et al. (\citeyear{Wharton2012}) inferred a recycling fraction of $f_{\rm rec} \geq 10^{-3}$.
Since the formation mechanism of MSPs is linked to accretion from a companion star, LMXBs have usually been considered the progenitors of MSPs. LMXBs are found to be $\sim 100$ times more abundant in the globular clusters than in the Galactic field \citep{Clark1975, Katz1975}. Based on this information Wharton et al. (\citeyear{Wharton2012}) estimated that the recycling fraction for globular cluster is $f_{\rm rec} \sim 0.1$.

To determine the number of MSPs that formed from neutron stars,
during the time spent in their host cluster, we proceed on with an analysis based on our current knowledge of the MSP luminosity function in the Galactic globular clusters \footnote{From this moment on, we implicitly assume that the NSC at the Galactic Centre formed from the assembly of "globular clusters". The clusters described with a King model then represent replica of the
globular clusters of the Milky Way, and MSPs inherit the properties they show in the Galactic globular clusters.  One cannot exclude that MSPs
form in the high density environment of the newly formed NSC.  Our analysis is confined to the populations of MSPs that have been dragged by the dynamical process studied in \cite{Tsatsi2017}. }. 

We proceed in steps: (i) we first use  the luminosity function to infer the number of MSPs in 
a selected  set of Galactic globular clusters; (ii) for each selected cluster  we record 
the total cluster luminosity and mass-to-light-ratio \citep[as measured in][]{Mclaughlin2005, Kimmig2015} and assign a total  stellar mass;
 (iii) using the value of the neutron star mass fraction inferred from the simulation by \cite{Sippel2013}, we compute the number of neutron stars in each cluster and then the 
  "recycling fraction", defined as the ratio of the number of active MSPs to the number of neutron stars that formed in the cluster; (iv)  we then associate to each Galactic globular cluster the "encounter rate" taken from Bahramian et al. (\citeyear{Bahramian2013}) which measures the number of close stellar encounters per unit time and is linked to the formation of LMXBs ad MSPs \citep{Pooley2003, Bahramian2013}. This quantity is measured as $\Gamma_{\rm c} \propto \int \rho^2/\sigma$ where $\rho$ is the stellar density and $\sigma$ is the total velocity dispersion. In this way we explore the dependence of the recycling fraction with the encounter rate.

As far as (i) is concerned  we adopt three fits to the MSP luminosity function \citep{Bagchi2011} that will be used  also in the analysis on
the detectability of the MSPs at the NSC.  Following  \cite{Bagchi2011}, we consider three possible log-normal model fits (all expressed in unit of mJy kpc$^2$: Model 1 with mean $\mu=-1.1$ and standard deviation $\sigma=0.9$, which is the same proposed by Faucher-Giguere and Kaspi (\citeyear{Faucher-Giguere2006}); Model 2 with mean $\mu=-0.61$ and standard deviation $\sigma=0.65$; and Model 3 with mean $\mu=-0.52$ and  standard deviation $\sigma=0.68$.

\begin{figure}
\begin{center}
\includegraphics[width=\columnwidth]{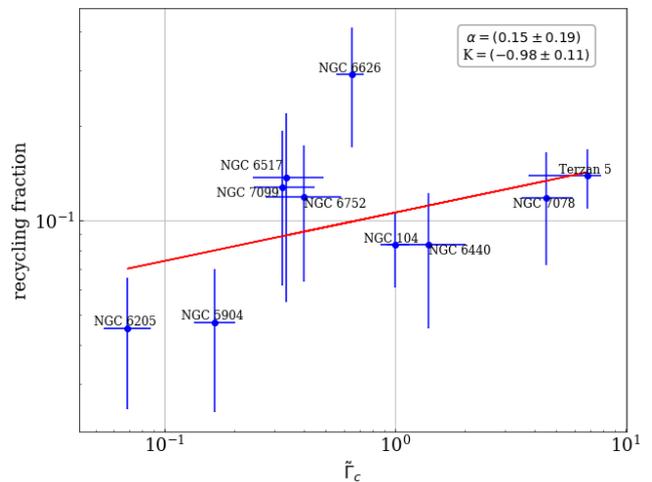}
\caption{Recycling fraction as a function of the normalised encounter rate for a number of Galactic globular clusters measured with Model 1 of \protect\cite{Bagchi2011}. The way these quantities are measured is described in the text. The parameters of the best power law fit are shown in the inset.
}\label{Recycling_fraction}
\end{center}
\end{figure}

Figure \ref{Recycling_fraction} shows the recycling fraction measured using Model 1 of Bagchi et al. (\citeyear{Bagchi2011}) as a function of the encounter rate, $\tilde \Gamma_{\rm c}$, normalised to the value obtained for NGC 104. The data is fitted with a power law of the form $\log(f_{\rm rec}(r)) = K + \alpha \log (\tilde \Gamma_{\rm c})$. The best fit parameters, obtained with a Markov Chain Monte Carlo (MCMC) algorithm, are $K=-0.98 \pm 0.11$ and $\alpha=0.16 \pm 0.19$ compatible with a straight line fit. 
The figure shows that over more than two order of magnitudes in $\tilde \Gamma_{\rm c}$ the value of $f_{\rm rec}$ varies weakly. The average value of $f_{\rm rec}$ of all clusters is 0.11. Using Model 2 and Model 3 of \cite{Bagchi2011} we see the same weak dependance on $\tilde \Gamma_{\rm c}$. The average value for Model 2 is 0.09 and for Model 3 is 0.06. Based on these results we can also consolidate the order of magnitude estimate of $f_{\rm{rec}} \sim0.1$ measured by \cite{Wharton2012} in a completely independent way. The value of the normalised encounter rate for the globular clusters simulated in this work is $\sim 20$. We assume that, for these values of the encounter rate, the recycling fraction remains the same  as that measured for the Galactic ones.

The encounter rate for the NSC can be measured using the density and velocity dispersion profiles with the same procedure used by \cite{Bahramian2013} for globular clusters. In this way we obtain a value, in normalised units, similar to the densest globular clusters, $\sim 10$ in the scale shown in Figure \ref{Recycling_fraction}. For this reason we assume that the recycling fraction is $\sim 0.1$ also for the NSC.
At the end of the simulations we randomly select the MSPs accordingly to the recycling fraction of the total NSs.  

\section{Observability of pulsars} \label{observability}

As anticipated in \S 1, no MSP has been discovered so far in the Galactic Centre area.  In this section we explore the observability of the population of MSPs inferred from the cluster-inspiral scenario, with more sensitive instruments than those available so far.
 
We have first to select the frequency band at which the surveys should be performed. While the steep radio spectrum would suggest to adopt the low frequencies (1.2-1.8 GHz) at which most of the searches in the Galactic plane have been run so far, the strong scattering of the Galactic Centre would advise to move to higher frequencies.  \citep{Macquart2015} suggested that the best compromise -- for the scattering scenario considered here -- is to operate the search experiments around a frequency of $\sim 8-9$ GHz. In view of that, we run our simulations for two reference cases of study. Case 1 assumes to utilise an instrument with characteristics similar to those originally proposed for the high frequency band (8-14 GHz) of the MeerKAT radio-telescope \citep{Booth2012}.  Case 2 appeals on a much higher instantaneous sensitivity (keeping the other instrumental parameters like those of Case 1), similar to that invoked for the band 5 (4.6-14 GHz) in the context of the SKA1-MID\footnote{www.skatelescope.org/wp-content/uploads/2012/07/SKA-TEL-SKO-DD-001-1$\mathunderscore$BaselineDesign1.pdf} baseline design. In both cases, we simulate surveys performed at a center frequency of 9 GHz with a bandwidth of 1 GHz.

The MeerKAT telescope is located in South Africa and is in the final phase of its construction \citep{Booth2012}. The SKA1-MID will be the first phase of the medium frequency part of the Square Kilometre Array (SKA) and will expand around the MeerKAT site \citep{Braun2015}. We note that the nominal capabilities of both the instruments at the $\sim 8-9$ GHz band are still under discussion. Hence, our simulations are not aimed at performing any detailed prediction about the MSP yields resulting from them. We used some general properties of the original design of those instruments to check if the order of magnitude of the MSPs discoveries is suitable to probe the formation history of the NSC at the Galactic Center.

We use the prescriptions described in \cite{Macquart2015} to measure the number of detectable MSPs.
A pulsar is detected if the pulsed emission is significantly stronger than the background noise, usually a threshold of SNR$=10$ is required, where SNR is the signal to noise ratio.
To measure the signal to noise ratio of a pulsar observation we use the radiometer equation. This equation states that, in the case of a narrow observing bandwidth, the SNR of a pulsar of flux density $S_{\nu}$, pulse period $P$ and width $W$ is \citep{Lorimer2006}:

\begin{equation}
{\rm{SNR}}= S_{\nu} \frac{G \sqrt{n_{\rm{p}} \Delta t \Delta \nu}}{\beta\, T_{\rm sys}} \sqrt{\frac{P-W}{P}},
\end{equation}
where $G$ is the telescope gain, $n_{\rm{p}}$ is the number of polarizations observed, $\Delta t$ is the total observation time, $\Delta \nu$ is the observing bandwidth and $T_{\rm sys}$ is the total system temperature. The factor $\sqrt{(P-W)/P}$ represents the fraction of power present in the pulse signal. In real surveys devoted to the search of new pulsars this performance in never fully reached. Imperfections caused by the digitisation of the signal and by the improper determination of pulsar properties can reduce the SNR.  The factor $\beta$ in the equation is called {\it correction fraction} and accounts for these effects. In the worst case scenario it can be as high as 2 and effectively half the observed SNR.

When $\Delta \nu$ is large, the assumption that $S_{\nu}$, $T_{\rm sys}$ and W are constant breaks down and we must consider their variations across the band. This can be done considering separately the total signal and the total noise received during the observation. The total signal measured can be written as:

\begin{equation}
S=\frac{1}{\beta} n_{\rm{p}} \Delta t \int_{\nu_1}^{\nu_2} S_{\nu} (\nu') d\nu',
\end{equation}
where $\nu_1$ and $\nu_2$ are the extreme of the observing band. For pulsars in the Galactic Centre, $S_{\nu}$ is usually written in the from $S_{\nu} (\nu)=L_{1.4}d_{\rm GC}^{-2} \left(\frac{\nu}{1.4 \, \rm{GHz}}\right)^{\alpha}$, where $L_{1.4}$ is the (pseudo-)luminosity of the pulsar at 1.4 GHz in units of mJy kpc$^2$, $d_{\rm GC}$ is the distance of the Galactic Centre in kpc and $\alpha$ is the spectral index.

The noise per polarization collected over a sampling time, $\delta t$ in a frequency channel $\delta \nu$ is $n(\nu)=\sqrt{\delta t \delta \nu}\, T_{\rm sys}(\nu) /G  \times\sqrt{W(\nu)/(P-W(\nu))}$. The total noise over all sampling times, frequencies and polarization channels adds in quadrature and therefore is:

\begin{equation}
N= \sqrt{ n_{\rm{p}}\Delta t \int_{\nu_1}^{\nu_2} \frac{T_{\rm sys}^2(\nu')}{G^2} \frac{W(\nu')}{P-W(\nu')} d\nu'}.
\end{equation}
Therefore, the SNR for the detection of a pulsar is calculated with the formula:

\begin{equation}
{\rm{SNR}}=\frac{1.62 \times 10^{-5} \sqrt{n_{\rm{p}} \Delta t} \int_{\nu_1}^{\nu_2} L_{1.4} \left(\frac{\nu'}{1.4 \, {\rm{GHz}}}\right)^{\alpha} (\nu') d\nu'}{\beta \sqrt{ \int_{\nu_1}^{\nu_2} \frac{T_{\rm sys}^2(\nu')}{G^2} \frac{W(\nu')}{P-W(\nu')} d\nu'}},
\end{equation} \label{radiometer_complete}
where we used $d_{\rm GC}= 7.86$ kpc \citep{Boehle2016}.

The telescope system temperature $T_{\rm sys}$ contains contributions from the sky and from the telescope receiver. The contribution from the sky can be divided in different parts: the emission from the the bright Galactic Centre region, the emission from the Earth's atmosphere and from the CMB and the noise of the receiver. Therefore it can be written in the form:

\begin{equation}
T_{\rm sys}(\nu)= T_{\rm GC}(\nu) + T_{\rm CMB} +T_{\rm atm} (\nu) + T_{\rm r}
\end{equation}
The contribution originating from the Galactic Centre can be modelled with the equation:

\begin{equation}
T_{\rm GC}(\nu) = T_0(\nu) \left( \frac{\theta_0^2}{\theta_0^2 +\theta_{\rm b}^2}\right),
\end{equation}
where $T_0(\nu)= 350 (\nu/2.7\rm{GHz})^{-2.7}$ K \citep{Reich1990}, $\theta_0=0^{\circ}.33$ \citep{Reich1990} and $\theta_b$ is the FWHM of the telescope beam, $\theta_b=1.22\lambda/d$, where $d$ is the diameter of the telescope. The contribution from the CMB is constant at $T_{\rm CMB} \sim 2.7$ K.

The atmospheric emission strongly depends on the telescope site, weather during the observations and on the elevation of the source. For an estimate of the atmospheric emission at the MeerKAT site we looked for the average archival temperature and humidity data for the site and, using past works \citep{Ajello1995,Ho2004,Ho2005}, we recovered the average value at the zenith at the observing frequency. Then we measured the temperature at the different elevations of the Galactic Centre from the telescope and averaged it over the observation. For the MeerKAT and SKA site at 9 GHz we obtained an average atmospheric noise of $\sim 5$ K. The temperature noise injected by the receiver is taken to be $\sim 10$ K, assuming performances similar to the current state-of-art systems. 

The observed width $W$ of the pulses plays an important role in mock surveys: in fact, when $W$  becomes of the order of the pulse period, the pulsar becomes undetectable even if the flux is very high. $W$ is influenced by different factors: the intrinsic width, $W_{\rm int}$, the time smearing due to scattering, $\tau_{\rm scat}$, the time smearing due to the dispersive nature of the ISM, $\tau_{\rm DM}$, as well as the sampling time, $\delta t$. The resulting width is the sum in quadature of these contributions:

\begin{equation}
W=\sqrt{ W_{\rm int}^2 + \tau_{\rm scat}^2 +\tau_{\rm DM}^2 + \delta t^2} 
\end{equation}

Following the indications of \citep{Dexter2017} for a Galactic Center region of  $\sim 50$ pc radius, 
the value of $\tau_{\rm scat}$ is taken to be the same as the one observed for the magnetar SGR J1745+29 \citep{Spitler2014}:

\begin{equation}
\tau_{\rm scat}= 1.3 \times \left( \frac{\nu}{1 \,\rm{GHz}}\right)^{-4} \rm s.
\end{equation}

The smearing effect due to the dispersive nature of the ISM follows the law:

\begin{equation}
\tau_{\rm DM}= 4.15 \times 10^{6}  \,\rm{DM}(\nu_1^{-2} - \nu_2^{-2})  \,\rm ms,
\end{equation}
where DM is the dispersion measure in units of pc cm$^{-3}$ and $\nu_1$ and $\nu_2$ are the extremes of the observing bands expressed in MHz. As a reference the magnetar SGR J1745+29 has a DM of $1778 \pm 3$ \citep{Eatough2013}. The corresponding time delay is $\sim 20$ ms at a central frequency of 9 GHz and a bandwidth of 1 GHz, enough to mask most millisecond pulsar. Luckily this effect can be compensated by splitting the data in narrow enough frequency channel and then adding a delay at each frequency channel of the opposite amount than that induced by the DM. This procedure (known as {\it dedispersion}) can almost completely correct for this effect, thus making the contribution of  $\tau_{\rm DM}$ negligible with respect to   $\tau_{\rm scat}.$ Therefore in our surveys we will assume $\tau_{\rm DM} \sim 0$. The adopted sampling time (see Table \ref{tabMeerkat}) is similar to what commonly used in current MSP searches.

We simulate surveys with integration times of 10 hours. While long integration times increase the sensibility for isolated MSPs, they might impede the discovery of binary MSPs. Usual binary pulsars search procedures lose efficiency when the orbital period of the binary is shorter than 10 times the total duration of the observation. Therefore binary MSPs with orbits shorter than 100 hours would be difficult to detect. Using the observed population of MSPs in globular clusters, \footnote{http://www.naic.edu/$\sim$pfreire/GCpsr.html} we see that $40 \%$ of all MSPs are in binaries with period shorter than this. Thus, these surveys would be sensible to only $60\%$ of the observed MSP population.

\begin{table}
\caption{Instrumental and observational parameters for simulated surveys. Case 1 refers to a pulsar search with parameters similar to those originally proposed for a Meerkat high frequency band survey. Case 2 refers to a survey with identical properties than Case 1, but for the significantly higher Gain of the telescope. This is meant to reflect the strong enhancement in the sensitivity provided by an experiment using a collecting area similar to that of SKA1-MID.}

\begin{center}
\begin{tabular}[t]{c c c}
\hline
{Parameter} & {Case 1} &{Case 2} \\
\hline
\hline
 Central frequency $\nu$ (GHz)& 9  & 9\\
Bandwidth $\Delta \nu$ (GHz) & 1 & 1  \\
Integration time $\Delta t$ (h) & 10 & 10 \\
Sampling time $t_{\rm{samp}} (\mu$s) & 40 & 40    \\
System temperature $T_{\rm{sys}}$ (K) & 32    & 32  \\
Gain $G$ (K/Jy) &   1.75  & 6.2\\
Max. baseline (m)&  1000  & 1000\\
FWHM &  8" (0.32 pc)   & 8" (0.32 pc) \\
\hline

\end{tabular}\\
\label{tabMeerkat}

\end{center}

\end{table}

In order to generate a population of synthetic MSPs in the Galactic Centre we assumed it to have the same properties as the ones observed in the Galactic disk and in the globular clusters. An important property is the beaming fraction, the fraction of pulsars whose emission crosses our line of sight and therefore visible. In the case of MSPs this fraction is very high, $0.5 - 0.9$ \citep{Kramer1998}.

Since the simulated survey will be performed at high frequency, $\sim 9$ GHz, while the luminosity distribution is measured at 1.4 GHz, the spectral index $\alpha$ also plays an important role. Studies directed on MSPs found that the spectral index has a mean value of 1.8 and a standard deviation of 0.6 \citep{Maron2000}. Similar values resulted from other studies \citep[e.g.][]{Toscano1998}. 

To simulate the other pulsar properties like period, period derivative, spin-down luminosity and pulse width, we used the measured parameters from the ATNF catalogue\footnote{http://www.atnf.csiro.au/people/pulsar/psrcat/} \citep{Manchester2005} for Galactic millisecond pulsars. The width of the pulse for MSP does not show strong variability as a function of frequency, therefore we opted to still use value measured at 1.4 GHz also for observations at very high frequencies.

The period derivative and spin-down luminosity do not enter directly in the process of radio detectability of pulsars, but they become important when considering the gamma ray luminosity. The gamma ray luminosity for pulsars is measured to be $L_{\gamma} \propto \dot E ^{1/2}$ \citep{Faucher-Giguere2010}, where $\dot E$ is the spin-down luminosity. If we consider only MSPs this law poorly represents the luminosities \citep{Abdo2013} and later works use the formula $L_{\gamma}= \eta \dot E$ with $\eta=0.2$ \citep{HooperLinden2016} or $\eta=0.05$ \citep{HooperMohlabeng2016}. As an approximate value we use $\eta=0.1$.

The last parameter we have not yet discussed is the luminosity distribution of the MSPs. Since the most used frequency for observing pulsars is 1.4 GHz, luminosities are usually scaled to this frequency to allow comparisons and are written in units of mJy kpc$^2$. As has already been discussed in the section \ref{Rec_fraction}, we will use the three log-normal models described in Bagchi et al. (\citeyear{Bagchi2011}).
These models give different predictions about the number and luminosity of the MSPs. While these models are all based on observations of pulsars in globular clusters, the difference between them comes from the different assumptions of low luminosity pulsars which are not observed. However, the low luminosity pulsars in the Galactic Centre would not be detected even with the next generation of radio telescopes. Using the formula \ref{radiometer_complete} we can find the lower limit luminosity of the MSPs that can be detected with a SNR higher than 10 in our simulated experiments. They range in the $4-13$ mJy kpc$^2$. Therefore in our survey we consider only the MSPs that are brighter than 1 mJy kpc$^2$. In the rest of the paper these MSPs will be referred to with the term "radio-bright MSPs". 
For this radio-bright branch of the luminosity function, all three models result in a comparable number of detections.

The final step is to find the number of radio-bright MSPs deposited in the NSCs. We proceed in the following steps. 
(i) First we calculate the fraction of radio-bright MSPs to the total number of MSPs using the luminosity function. The desired fraction will be the value of the integral of the luminosity function with luminosities above the threshold considered (the luminosity function is normalised to one). (ii) Repeating the analysis of section \ref{Rec_fraction} we determine the fraction of radio-bright MSPs to neutron stars in the Galactic globular clusters. The average value for this fraction is $\sim 0.013$. This fraction appears to be independent of the luminosity function model used  (as it is expected from the fact that all models give the same number of radio-bright MSPs). The number of radio-bright MSPs is roughly a tenth of all MSPs. (iii) We use this fraction to randomly extract radio-bright MSPs from the population of neutron stars in the NSC at the end of the simulation.

\section{Results} \label{results}

In order to check for the observability of the population of MSPs, the results of our cluster-inspiral scenario need to be projected along a realistic line of sight. We choose to project along the line of sight that maximises the rotation of the NSC in order to reproduce the observed rotation \citep{Tsatsi2017}. The projected distribution maintains the same distribution as is observed in Figure \ref{distribution_together} but the peak moves to $\sim 2$ pc.

\begin{figure}
\begin{center}
\includegraphics[width=\columnwidth]{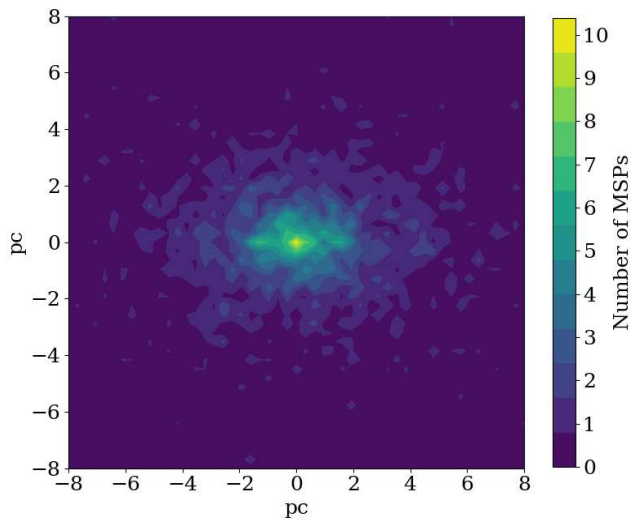}
\caption{Colour-scale map of the distribution of the MSPs at  the Galactic Centre. The map is built on a grid of pixels size $\sim 0.32$ pc, which corresponds to the telescope beam for Case 1 and Case 2.}\label{NS_map}
\end{center}
\end{figure}

We divide the central parsecs in bins of width equal to the beam of the telescope, and count the number of radio-bright MSPs in each bin. The width of the beam -- which will be obtained from the combination of the voltages collected at the various antennas of the arrays by using a beam-forming procedure  -- is assumed to match the diffraction limited resolving power; in particular (see Table 1) that implies a width of $8"$ which in turn corresponds to $\sim 0.32$ pc at the distance of the Galactic Centre \citep[assumed to be $7.86$ kpc][] {Boehle2016}. The map of the intrinsic distribution of the MSPs is displayed in Figure \ref{NS_map}, using a bin size equal to the beam width of the telescope.

\subsection{Previous surveys}

The most sensitive pulsar survey of the Galactic Centre as of today was performed at Green Bank Telescope at a frequency of 14.8 GHz \citep{Macquart2010}. The telescope has an average diameter of $\sim 100$ m so the beam of the observation will have a width of $50"$, corresponding to $\sim 2$ pc at the Galactic Centre.
Using the recipes described in the previous section, repeated simulations of the aforementioned survey results in the average detection of $\sim 1$ MSPs. In this case, according to Poissonian statistics,  the probability of not observing any MSPs is $\sim 35\%$. The null result of the experiment does not contradict the predictions of our simulations.

\subsection{Case 1- MeerKAT-like survey} \label{results_meer}

\begin{figure}
\begin{center}
\includegraphics[width=\columnwidth]{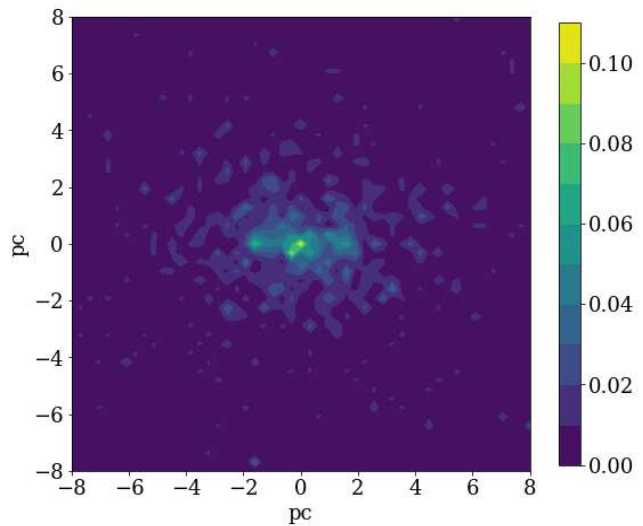}
\caption{Color-scale map reporting the fractional probability of a MSPs at the Galactic Centre to be detected in a Case 1 experiment, averaged over one hundred simulations.
}\label{MSP_observed_Meerkat}
\end{center}
\end{figure}

In Case 1, we simulate the observable sample considering an observation time  of 10 hours (see Table 1), in agreement with the observability of the Galactic Centre above the minimum altitude of 15$^{\circ}$.  For each bin we run the code to determine how many of the radio-bright MSPs are detectable with a MeerKAT-like radio telescope. The result of the simulations is reported in Figure \ref{MSP_observed_Meerkat}. There is no single pixel where the detection probability is higher than about $10\%.$ We can potentially detect only $\sim 1$ MSP within 1 pc,  up to $\sim 6$ in a region with radius of 5 pc, and up to $\sim 15$ if the search is performed over the entire NSC. In the pessimistic case scenario, where the correction fraction $\beta$ (see \S \ref{observability}) is $\sim 2$ we obtain a total number of detections of order 10.

\subsection{Case 2- SKA1-MID like survey} \label{results_ska}

\begin{figure}
\begin{center}
\includegraphics[width=\columnwidth]{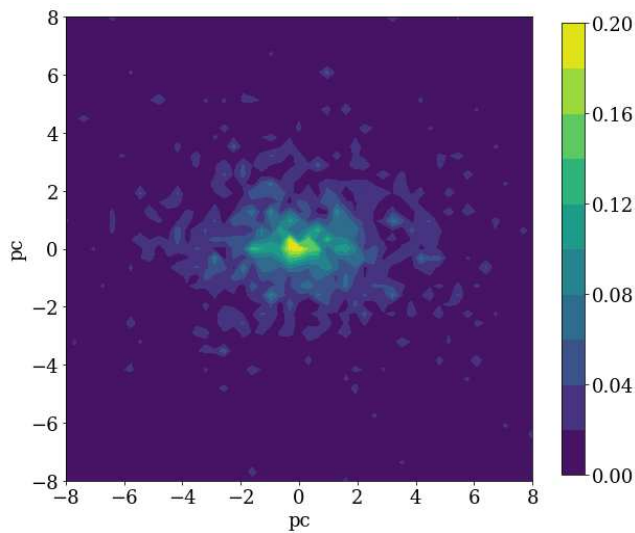}
\caption{Color-scale map reporting the fractional probability of a MSPs at the Galactic Centre to be detected in a Case 2 experiment, averaged over one hundred simulations.
}\label{MSP_observed_ska}
\end{center}
\end{figure}

In the case of a SKA1-MID-like survey we simulate observations of same duration as for Case 1. The result of the simulations is reported in Figure \ref{MSP_observed_ska}.  Thanks to the increased collecting area of this telescope the number of detections is significantly larger. Within 1 pc from the Centre, we still detect only $\sim 2$ MSPs, but the sample increases to $\sim 20$  in a radius of 5 pc and to $\sim 50$ if we consider the entire NSC. In the pessimistic case scenario, where the correction fraction $\beta$ is $\sim 2$ we obtain a total number of detections of $\sim 30$.

\section{Discussion} \label{discussion}

In our cluster-inspiral scenario,  about 2700 MSPs are present in the NSC. 
Only a fraction, $\sim 5\%$ are found in the central parsec, and the bulk extends over a region of $\sim 20$ pc. 
The aggregation and tidal disruption of globular clusters is a mechanism able to disseminate
     MSPs over the entire NSC. 
     The mass segregation of neutron stars in the NSC is not included in the simulations during after the assembly of the NSC.  We verified that this effect is to first approximation negligible since their mass segregation timescale exceed the Hubble time \citep{Freitag2006,Merritt2010}.

 Not all MSPs that form in the globular clusters, released in the NSC, will be visible with future radio telescopes. In Case 1 (MeerKAT-like telescope), the detectable MSPs are of order unity within 1 pc, and of order dozen over a spherical volume of $\sim 20$ pc radius.  In Case 2  (SKA1-MID-like telescope) we expect a significant improvement of a factor $\sim 3$ in the yields. 
 
\cite{Wharton2012}  considered an in-situ formation scenario for the MSPs at the Galactic Centre, and estimated a number of about  $1000$ MSPs within the central parsec. Since we don't have enough information to probe the spatial distribution of this population, we only focused on the total number in the central parsec.
By simulating a survey, we verified that the Green Bank Telescope would have detected $\sim 2$ MSPs, consistent with the null result
of the recent survey by \cite{Macquart2010}. 
A MeerKAT-like telescope and a SKA1-MID-like telescope would detect   $\sim 5$ and $\sim 20$ MSPs, respectively in the central parsec.
These are much numerous populations than those implied by the cluster-inspiral scenario and a statistically sound discrimination among the {\it in-situ}  and the {\it cluster-inspiral} hypotheses could be obtained, at least for an experiment of the scale of SKA1-MID (Case 2).

Our estimates depend on the extent of the interstellar scattering at the Galactic Centre, which is the main responsible for smearing the signal from MSPs. In this work, we considered  the scattering to be comparable to that inferred from the radio pulses of the magnetar SGR J1745-29. We note that most of the discoverable MSPs in our scenario are indeed outside the central parsec and extend up to $\sim 20$ pc. These distances are comparable with those of some ordinary pulsars discovered in the Galactic Centre region which show weak scattering of the same entity of the magnetar. Therefore the assumption of a not destroying effect of the scattering at 8-9 GHz is plausible at the positions where we expect to make most of the detections. 

However, if the scattering in the central region will turn out to be stronger and close to the values reported in \cite{Lazio1998}, the detections of MSPs will be harder. In this case, effective surveys would have to be conducted at much higher frequencies in order to make the scattering unimportant. \cite{Macquart2015} indicated that $\sim 25$ GHz might be a good choice when finding for a trade-off among the aim of reducing the scattering smearing and the need of not missing the MSPs due to the weakening of their signal. 

\subsection{Gamma-ray emission}

We now discuss the possible contribution of the derived population of MSPs to the gamma-ray emission observed by the {\it Fermi} satellite in the Galactic centre region, limiting the analysis
to the MSPs confined within the NSC resulting from the assembly of in-spiralling globular clusters.

We estimate that a population of $\sim 2700$ MSPs are present in the inner 20 pc of the Galaxy. Some of these MSPs could have been seen as point sources in the \emph{Fermi} map. An approximate estimate of the point source threshold in this region is $5 \times 10^{34}$ erg s$^{-1}$ \citep{Haggard2017}. Averaging the result of our simulation over 100 trials we find that, in average, $5$ MSPs will exceed this threshold.  The null result of current observations are compatible with the predictions only at the 3 sigma level. 

However, we note that the majority of the MSPs are located outside the central parsecs of the Galactic Center. In this region the MSPs will mostly evolve in isolation with very small probabilities of new events of recycling. The MSPs will therefore lose rotational energy through magnetic dipole breaking. The gamma-ray luminosity, as described in $\S$ \ref{observability}, is linked to the spin-down luminosity and will decrease significantly over the timespan of the simulation. The number of MSPs with luminosities above the threshold will decrease accordingly. 

In light of these considerations we limit our discussion of the gamma-ray excess to the central 5 pc of the Galaxy. The total gamma-ray luminosity of the excess is $(2.0 \pm 0.4)\times 10^{37}$ erg s$^{-1},$ calculated within $\sim 1$ kpc from the centre \citep{Calore2016,Haggard2017}. If  we rescale the gamma-ray luminosity to account for the emission within  the inner 5 pc according to the spatial profile of the excess \citep{Calore2015}, we find a value of $\sim1.5 \times 10^{36}$ erg s$^{-1}.$ 
Since the average gamma-ray luminosity emitted by a MSPs in our model is $\sim 1.4 \times 10^{33}$ erg s$^{-1},$ the number of required MSPs is $\sim 1000$. In this region we find $\sim 1000$ MSPs. We can therefore explain the observed excess in the inner 5 pc of the Galaxy. 
We note that similar results have been found by \citeauthor{Arca-Sedda2017c} (\citeyear[][private communication]{Arca-Sedda2017c}) using different simulations and different methods to estimate the final MSP population. In this work the authors conclude as well that the deposition of MSPs from inspiralling globular clusters can explain the observed central GeV excess.

\section{Conclusions}

In this work we considered the cluster-inspiral scenario for the formation of the NSC at the Galactic Centre, based on the simulations described by \cite{Perets2014} and \cite{Tsatsi2017}.
We showed that  this scenario predicts the existence of a  population of MSPs  resulting from the in-spiral and dismemberment of globular clusters
hosting recycled neutron stars.  The neutron star population has been inferred from a synthesis model by \cite{Sippel2013} that accounts for stellar evolution and mass segregation.  The key finding is that MSPs are distributed over the entire NSC with no significant clustering within the central parsec. 

If the effect of the scattering is similar to that shown by the magnetar SGR J1745-29, which is so far the closest
pulsar to the Galactic Centre,  future experiments with reasonable observing parameters and operating at $8-9$ GHz will be able to detect a significant number of these MSPs. By contrast, if MSPs form in the centre-most regions around the massive black hole via an in-situ mechanism, a detectable population of MSPs could arise in the central parsec. If a considerable number of MSPs will be found outside the central few parsecs of our Galaxy, this will provide evidence in favour of  the cluster-inspiral scenario for the formation of the NSC.

The in-situ and cluster-inspiral processes are not mutually exclusive and the current models are still lacking of important details to be worked out. On a theoretical side, the globular cluster model that is used to sample the neutron star populations, should be improved for future similar studies. In addition, the $N$-body simulations of the cluster-inspiral should include stellar evolution extended over $\sim 12 $ Gyrs as well as the effects of secular dynamics on the neutron star population that the NSC inherits from the globular clusters. On an observational side, in the case of unfavourable conditions due to the interstellar scattering, searches at higher frequencies than those explored here will be necessary to detect MSPs in the NSC at the Galactic Centre.

\section*{Acknowledgements} 

With the support of the Italian Ministry of Foreign Affairs and International Cooperation, Directorate General for the Country Promotion (Bilateral Grant Agreement ZA14GR02 - Mapping the Universe on the Pathway to SKA).
AMB and ACS acknowledge support by Sonderforschungsbereich SFB 881 `The Milky Way System' (subproject A7 and A8) of the German Research Foundation (DFG).



\bibliographystyle{mnras}
\bibliography{pulsar_Galactic_Center}







\bsp	
\label{lastpage}
\end{document}